\documentclass[10pt, conference, letterpaper]{IEEEtran}
\IEEEoverridecommandlockouts
\usepackage[utf8]{inputenc}
\usepackage{tabularx,tabulary, adjustbox}
\usepackage{xcolor,tcolorbox,colortbl}
\usepackage{soul}
\usepackage[misc,clock,geometry]{ifsym}
\usepackage{amsthm,amssymb,graphicx,multirow,amsmath,color,amsfonts}
\usepackage{caption, subcaption, cite, epstopdf, comment, mathtools}
\usepackage{pifont, mdsymbol}
\usepackage{breakurl}
\usepackage{array}

\newcolumntype{C}[1]{>{\centering\arraybackslash}m{#1}}
\newcolumntype{Z}{>{\raggedright\let\newline\\\arraybackslash\hspace{0pt}}X}

\newcommand{\cut}[1]{}

\newcommand{\subparagraph}{}
\usepackage{titlesec}    

\newcommand{\ts}{\textsuperscript}

\long\def\/*#1*/{}

\newcommand{\confname}{IEEE International Conference on Computer Communications (INFOCOM)}
\newcommand{\conflocation}{Vancouver, Canada}
\newcommand{\confdate}{May 2024}

\usepackage{fancyhdr}
\fancypagestyle{firstpagestyle}{
\fancyhf{}
\lhead{\small Author's accepted version for \confname, \conflocation, \confdate}

\lfoot{\small © 2024 IEEE.  Personal use of this material is permitted.  Permission from IEEE must be obtained for all other uses, in any current or future media, including reprinting/republishing this material for advertising or promotional purposes, creating new collective works, for resale or redistribution to servers or lists, or reuse of any copyrighted component of this work in other works}
}

\def\BibTeX{{\rm B\kern-.05em{\sc i\kern-.025em b}\kern-.08em
    T\kern-.1667em\lower.7ex\hbox{E}\kern-.125emX}}

\begin{document}
\include{notation}

%\titleformat{\subsubsection}
%  {\normalfont\normalsize\textbf}{\thesubsubsection.}{1em}{}

\graphicspath{{./Figures/}}

\title{5G-WAVE: A Core Network Framework with Decentralized Authorization for Network Slices}

\author{
\IEEEauthorblockN{Pragya Sharma\IEEEauthorrefmark{1},
Tolga Atalay\IEEEauthorrefmark{1},
Hans Andrew Gibbs\IEEEauthorrefmark{2},
Dragoslav Stojadinovic\IEEEauthorrefmark{2},\\
Angelos Stavrou\IEEEauthorrefmark{1}\IEEEauthorrefmark{2},
Haining Wang\IEEEauthorrefmark{1}}
\IEEEauthorblockA{\IEEEauthorrefmark{1}Bradley Department of Electrical and Computer Engineering, Virginia Tech, USA\\
\IEEEauthorrefmark{2}Kryptowire LLC, McLean, VA, USA\\
Email: \IEEEauthorrefmark{1}\{pragyasharma, tolgaoa, angelos, hnw\}@vt.edu, \IEEEauthorrefmark{2}\{hgibbs, dstojadinovic, angelos\}@kryptowire.com}

%\thanks{This material is based on research sponsored by the U.S. Defense Advanced Research Projects Agency (DARPA) under agreement number HR001120C0155. The views, opinions, and/or findings contained in this article are those of the author(s) and should not be interpreted as representing the official views or policies, either expressed or implied, of the Defense Advanced Research Projects Agency or the Department of Defense.
%}
}

\maketitle
\thispagestyle{firstpagestyle}

% \fancyfoot[L]{© 2024 IEEE.  Personal use of this material is permitted.  Permission from IEEE must be obtained for all other uses, in any current or future media, including reprinting/republishing this material for advertising or promotional purposes, creating new collective works, for resale or redistribution to servers or lists, or reuse of any copyrighted component of this work in other works}

% \lfoot{© 2024 IEEE.  Personal use of this material is permitted.  Permission from IEEE must be obtained for all other uses, in any current or future media, including reprinting/republishing this material for advertising or promotional purposes, creating new collective works, for resale or redistribution to servers or lists, or reuse of any copyrighted component of this work in other works}

% \setlength{\footheight}{9pt}
%9 lines to not exceed limit

\begin{abstract}

5G mobile networks leverage Network Function Virtualization (NFV) to offer services in the form of network slices. Each network slice is a logically isolated fragment constructed by service chaining a set of Virtual Network Functions (VNFs). The Network Repository Function (NRF) acts as a central OpenAuthorization (OAuth) 2.0 server to secure inter-VNF communications resulting in a single point of failure. Thus, we propose 5G-WAVE, a decentralized authorization framework for the 5G core by leveraging the WAVE framework and integrating it into the OpenAirInterface (OAI) 5G core. Our design relies on Side-Car Proxies (SCPs) deployed alongside individual VNFs, allowing point-to-point authorization. Each SCP acts as a WAVE engine to create entities and attestations and verify incoming service requests. We measure the authorization latency overhead for VNF registration, 5G Authentication and Key Agreement (AKA), and data session setup and observe that WAVE verification introduces 155ms overhead to HTTP transactions for decentralizing authorization. Additionally, we evaluate the scalability of 5G-WAVE by instantiating more network slices to observe 1.4x increase in latency with 10x growth in network size. We also discuss how 5G-WAVE can significantly reduce the 5G attack surface without using OAuth 2.0 while addressing several key issues of 5G standardization.

\end{abstract}
%the abstract is already uploaded

%%%%%%%%%%%%%%%%%%%%%%%%%%%%%%%%%
% \section{Introduction} \label{sec:intro}
%\input{introduction.tex}
\section{Introduction}

The deployment of Fifth-Generation (5G) mobile networks is an ongoing process that is gaining momentum with each passing year. As part of their service offerings, 5G mobile networks have to meet the strict Quality of Service (QoS) requirements for a diverse set of use cases. %These include enhanced Mobile Broadband (eMBB), Ultra Reliable Low Latency Communication (URLLC), Massive Internet of Things (mIoT), and Vehicular to Everything (V2X) communications~\cite{3gpp23501}. 
To achieve such versatility, 5G services are offered as logically isolated end-to-end fragments known as network slices. This approach enables 5G operators to create independent network instances that can be tailored for different sets of requirements. %such as various bandwidth, latency, and QoS guarantees to deliver an end-to-end service tailored for specific applications. %For example, a network slice for autonomous vehicles would require ultra-low latency and high reliability, while a slice for a video streaming services would need high bandwidth and throughput.

Compared to Long Term Evolution (LTE), delivery of services through network slicing relies on Network Function Virtualization (NFV) for increased flexibility. Leveraging NFV, network slices are constructed by 
%around a service-based architecture (SBA) constructed through the
service chaining a set of Virtual Network Functions (VNFs) hosted on Commercial Off-The-Shelf (COTS) hardware. This allows operators to scale their deployments more efficiently and reduce costs by replacing dedicated service hardware. The network slice VNFs communicate with one another through HTTP transactions within a Service-Based Architecture (SBA)~\cite{3gpp23501}. 
A key feature of the 5G core SBA is the use of standardized Application Programming Interfaces (APIs) to enable interoperability between alternative implementations of the 5G core VNFs. The individual HTTP transactions between the VNFs are secured using an OpenAuthorization (OAuth) 2.0 server that distributes access tokens for moderating access requests~\cite{3gpp33501}. 
%In this paper, we propose replacing this token-based approach with a decentralized authorization framework. 

% To illustrate the network slicing layout in the 5G core SBA, a tentative deployment is shown in Figure~\ref{fig:sliceconsbas}.
% As the key enabler of inter-VNF authorization, the Network Repository Function (NRF)~\cite{3gpp29510} maintains the metadata profiles of the VNFs in a given administrative domain. The NRF orchestrates the application-layer communication between VNFs for network slice construction through a series of mutual discovery operations that establish connections between service providers and requesters.

According to the 3rd Generation Partnership Project (3GPP)~\cite{3gpp29510, 3gpp33501} standardizations, the Network Repository Function (NRF) maintains the metadata profiles of all the VNFs and supports mutual discovery operations that establish connections between service providers and requesters in a given administrative domain. 3GPP also defines NRF as an OAuth 2.0 server that distributes access tokens to enable inter-VNF authorization for the secure consumption of services. 

\textbf{Motivation.} While this fundamental approach to service discovery and authorization fulfills the basic requirements for security, it promotes a rigid framework for network slice construction where VNFs rely on a central trusted entity. In a scenario where the NRF is compromised, an adversary can issue malicious tokens or tamper service contracts for the provider VNFs. Moreover, the 5G SBA is a distributed ecosystem deployed as microservices, instead of monolithic VNFs. Thus, the dependency on a central entity for authorization can lead to performance bottlenecks in control plane signaling.

To address this security gap, we propose an integrated platform called \textbf{5G-WAVE} for network slice construction with decentralized authorization among the 5G core VNFs. WAVE~\cite{Andersen2019} is a decentralized authorization and verification framework which enables transitive delegation of access among entities without relying on a central trust authority.
%
%an authorization and verification engine for distributed systems. 
%
%It provides transitive delegation of access among entities without relying on a central trust authority. 
%Furthermore, WAVE borrows from existing decentralized Trust Management (TM) systems and improves them with additional security guarantees. The delegations and revocations of permissions are cryptographically enforced offering confidentiality and invisibility to untrusted parties. 
%
It allows authorization between \textit{`WAVE entities'} by creating contracts called \textit{`attestations' }for resource access. 
%T
When service requests arrive, WAVE uses
these attestations to verify their permissions. %access of service requests.

%produced during handling HTTP requests for verification before the request is completed. 
%When handling HTTP requests, WAVE consults attestations that were previously constructed and signed during the permission-granting process. These attestations are used to verify authorization before the completion of the request.

\begin{comment}
\begin{figure}[t]
    \centering
    \includegraphics[width=\columnwidth,trim={8.6cm 7.7cm 10.6cm 7.2cm},clip]{figures/sliceconsbas.pdf}
     \caption{Network slicing overview with NRF serving as a metadata storage and authorization server for multiple slices}
     \label{fig:sliceconsbas}
\end{figure}
\end{comment}

%Initially designed for distributed systems like IoT devices in a built environment \cite{Anderson2018, AndersonThesis}, WAVE can be used as a general-purpose decentralized authorization framework. 
%It allows authorization between \textit{`WAVE entities'} by creating contracts called \textit{`attestations' }for resource access. 
%These attestations are produced during HTTP requests where access is verified before the request is completed. 
% In our proposed framework, WAVE is seamlessly integrated into the 5G ecosystem through the design of a custom authorization reverse proxy, deployed adjacent to each core VNF. 

To integrate WAVE seamlessly into the 5G ecosystem, we propose a microservice-based augmentation to a Kubernetes-based 5G core deployment. The WAVE entities are deployed adjacent to each VNF as Side-Car Proxies (SCP)~\cite{EdpriceMsft2022Jun}. Using the SCPs, an indirect communication service mesh~\cite{3gpp29500} is created around the 5G core to offload the authorization functionality from the 5G core VNF to the WAVE infrastructure. The SCPs serve as WAVE entities granting attestations for resource access, thereby removing the reliance on a central token-issuing server for authorization. To facilitate implementation and integration, we use the WAVE SCP (wSCP) in conjunction with a redirection SCP (rSCP) to create an interception/verification pipeline for the 5G core service requests. 
If the service request is verified successfully, it is forwarded to its original destination, i.e., the target 5G core VNF. In our evaluations, we measure the latency overhead introduced to the 5G operations because of integrating WAVE into the core network. The results show that an authorization latency of 155ms is introduced to individual HTTP transactions. As we scale the slices, we observe a 1.4x increase in authorization latency with 10x network growth. On the other hand, the 5G-WAVE integration addresses several key issues from the 3GPP network slicing security enhancement efforts. Furthermore, by proposing an alternative to OAuth 2.0, multiple attack vectors that leverage the OAuth threat surface are eliminated.

%to authorize each other without going through the NRF for access tokens. Similarly, they can provide a proof of their permissions during service discovery directly to the requesting client.% which can improve system efficiency.

%In the 5GC, multiple vendors who are unknown to each other deploy their VNFs in a virtualized public cloud. In addition to intra-slice authorization among VNFs, WAVE can offer privacy protection in such cases by restricting visibility of permissions to relevant vendors only. Network slices with VNFs in 5G can be thought of as domains of hierarchy of resources and WAVE can facilitate inter-slice communication as it supports cross administrative domain interaction.

\begin{figure}[t]
    \centering
    \includegraphics[width=\columnwidth,trim={3.8cm 1.5cm 2.2cm 0.5cm},clip]{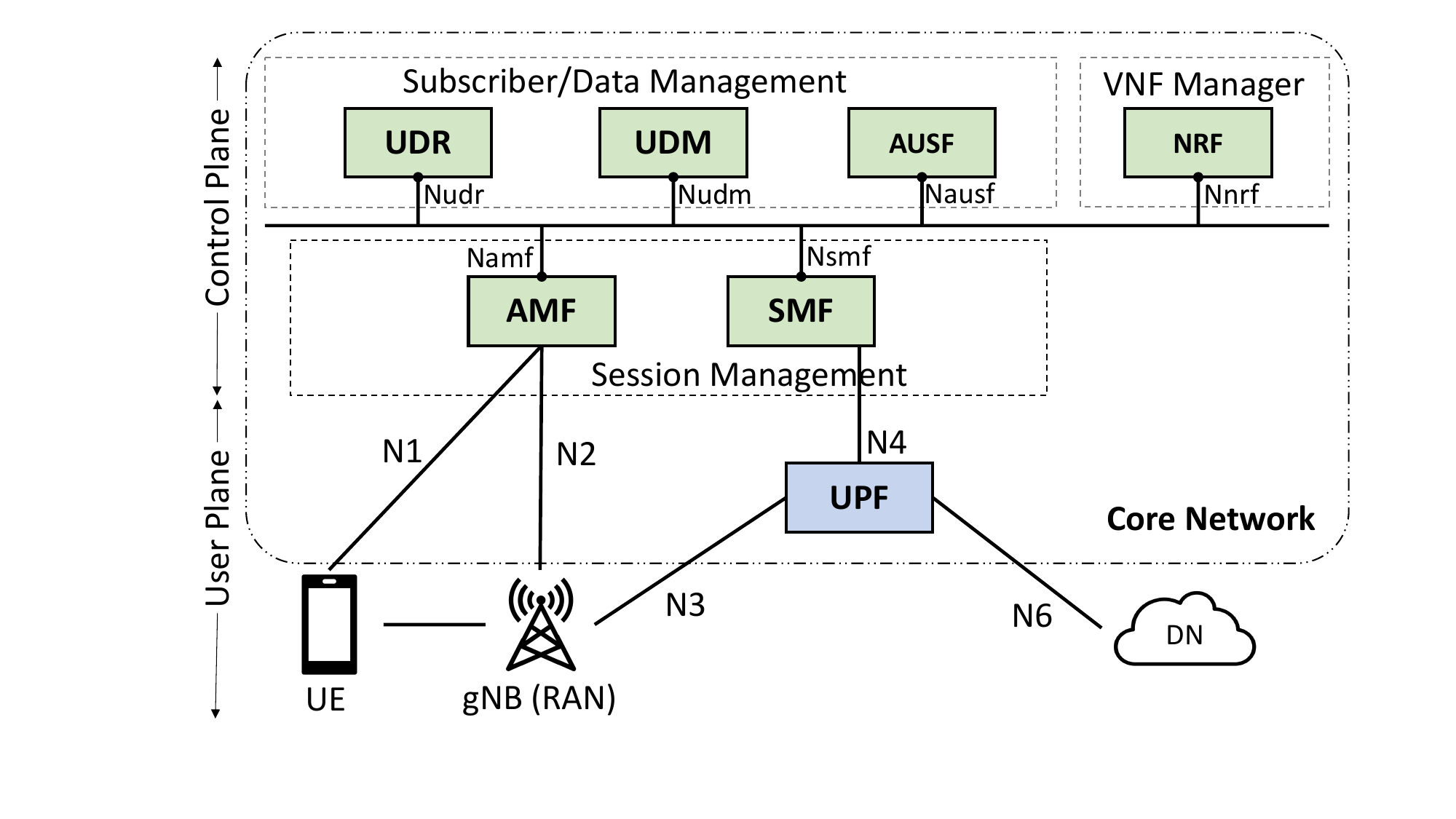}
     \caption{5G core service-based architecture}
     \label{fig:5gsa}
\end{figure}

The main contributions of this work are summarized below.
\begin{itemize}
    \item We propose the 5G-WAVE integrated platform  for decentralized inter-VNF authorization. 
   \item We implement the authorization flow using the OpenAirInterface (OAI)~\cite{OpenAirI28online} 5G core and test it using gNBSIM~\cite{RohanGnb17online}. An interception/verification pipeline is constructed using two SCPs to facilitate the seamless integration of WAVE into the 5G core.
    \item We measure the latency overhead of the proposed platform in comparison to native 5G core with single and multiple network slices.
    \item A security analysis is conducted to demonstrate how the proposed 5G-WAVE integration addresses key security issues from the 3GPP standardization as well as the OAuth attack surface.
\end{itemize}

%The outline of this paper is as follows. In Section~\ref{sec:motiv}, we provide the motivation behind this research. Section~\ref{sec:backg} presents the background information on the 5G system architecture, standardized 3GPP authorization framework and WAVE fundamentals. We present our proposed 5G-WAVE integrated platform design along with the intra/inter-pod message flows and implementation details in Section~\ref{sec:framework}. In Section~\ref{sec:experiment}, we describe our testbed used to conduct scalability experiments for the integrated 5G-WAVE platform. The evaluation is conducted in Section~\ref{sec:perfeval} followed by a discussion in Section~\ref{sec:discussion}. We explore the related work on decentralized 5G authorization in Section~\ref{sec:relwork}. Finally, the paper is concluded with Section~\ref{sec:conc}.
 \label{sec:intro}
%%%%%%%%%%%%%%%%%%%%%%%%%%%%%%%%%
%\input{motivation.tex} \label{sec:motiv}
%%%%%%%%%%%%%%%%%%%%%%%%%%%%%%%%%
\section{Background} \label{sec:background}
\subsection{5G Core Overview}

The 5G core has a SBA comprised of a set of interconnected VNFs as shown in Figure~\ref{fig:5gsa}. Each VNF needs to be properly authorized to access the resources of another~\cite{3gpp23501}. The key network functions of the 5G core are summarized as follows: 
%
% 5G core architecture is defined as a service-based architecture (SBA) with services offered by VNFs \cite{3gpp23501}. The SBA provides higher flexibility and better modularization of the 5G system for different network slices. In a 5G network deployment, the operators identify the requirements of the customer and assign a set of dedicated or shared VNFs better known as a network slice instance (NSI). Some of the key VNFs instantiated in a NSI of the 5GC network are:
%
    % \item 
    1) Network Repository Function (NRF) serves as a centralized VNF profile repository and facilitates mutual discovery and authorization. 
    % \item 
    2) Unified Data Repository (UDR) serves as a centralized repository for subscriber data.  
    % \item 
    3) Unified Data Management (UDM) manages the subscriber context data and authentication credentials.
    % \item  
    4) Authentication Server Function (AUSF) terminates the 5G Authentication and Key Agreement (AKA) service chain in the 5G home network.
    % \item 
    5) Access and Mobility Management Function (AMF) manages the subscriber access and mobility within the network.
    % \item 
    6) Session Management Function (SMF) manages the subscriber data session within the network. 
    % \item 
    7) User Plane Function (UPF) handles the user traffic by forwarding it between the user and Data Nodes (DNs). 
%The core network is divided into a control plane and a user plane (Figure \ref{fig:5gsa}), where NRF, UDR, UDM, AUSF, AMF and SMF form the control plane and the UPF is part of the user plane. UPF connects the core network with the user equipment (UE), radio access network (RAN) and the data network (DN). 5G supports multitude of applications by leveraging the SBA to support multiple virtual networks in form of `slices' that operate on the same physical hardware. An end-to-end logically isolated instance of a set of control plance and user plane VNFs is called a Network Slice Instance (NSI).
% 3GPP has defined a standardized method of VNF discovery and authorization within the 5G architecture~\cite{3gpp29510}. An overview of this process is shown in Figure~\ref{fig:5gdiscov} for an AMF seeking to discover an SMF.

\begin{figure}[t]
    \centering
    \includegraphics[width=0.98\columnwidth, trim={1.7cm 11.6cm 1.8cm 1.2cm},clip]{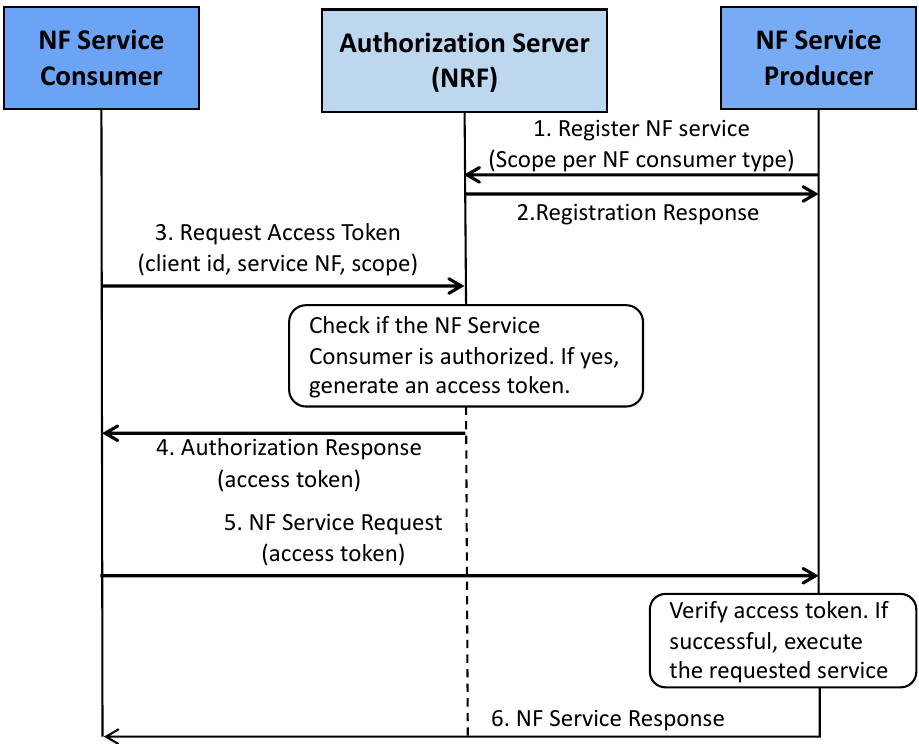}
     \caption{NRF as OAuth2.0 server for inter-VNF authorization}
     \label{fig:5goauth}
\end{figure}

\subsection{5G Core Authorization - Current Standard}

3GPP has standardized OAuth 2.0~\cite{3gpp33501,3gpp29510} as the default authorization mechanism between VNFs. It denotes the NRF as an OAuth 2.0 server. Any other VNF Service Consumer (aka service requester) acts as an OAuth 2.0 client and the VNF Service Producer (aka service provider) acts as an OAuth 2.0 resource server. As illustrated in Figure~\ref{fig:5goauth}, the service producer registers with the NRF as an OAuth 2.0 resource server along with the scope of authorization for service consumers. When the service consumer needs to access the services of the producer, it sends an authorization request to the NRF along with the requested scope and relevant access data of the target provider. The NRF checks whether the consumer is authorized to access the requested service. If it is authorized, then the NRF generates an access token with the appropriate scope and forwards it to the consumer as a response. After authorization, the consumer sends a service request to the provider with the access token. The service provider validates the received access token and responds with the requested resource. 

% \begin{figure}[t]
%     \centering
%     \includegraphics[width=\columnwidth, trim={5cm 0cm 4.5cm 0cm},clip]{figures/5goauth.pdf}
%      \caption{NRF as OAuth2.0 server for authorization and verification of VNFs}
%      \label{fig:5goauth}
% \end{figure}

\begin{comment}
    \begin{figure}[h]
    \centering
    \includegraphics[width=\columnwidth,trim={8.3cm 2.5cm 8.3cm 1.2cm},clip]{figures/5gdiscflow.pdf}
     \caption{NRF service discovery and authorization example}
     \label{fig:5gdiscov}
\end{figure}
\end{comment}

\begin{comment}
\begin{figure}[h]
    \centering
    \includegraphics[width=\columnwidth,trim={0cm 10.2cm 0cm 9.8cm},clip]{figures/5gdiscov.pdf}
     \caption{NRF service discovery and authorization example}
     \label{fig:5gdiscov}
\end{figure}
\end{comment}

\begin{comment}

The AMF will start by (1) sending a discovery request detailing the target SMF it wished to discovery depending on the network slice being constructed. NRF will (2) search the internal database and (3) respond with a list of desired SMF profiles. After selecting an SMF, the AMF will (4) request and access token for it and (5) receive one from the NRF. The AMF will (6) attach this token in the header of future API calls to the target SMF. 
    
\end{comment}

\subsection{Side-car Proxy and Service Mesh}
\begin{figure}[t]
    \centering
    \includegraphics[width=\columnwidth,trim={6.1cm 5.8cm 12.5cm 6.6cm},clip]{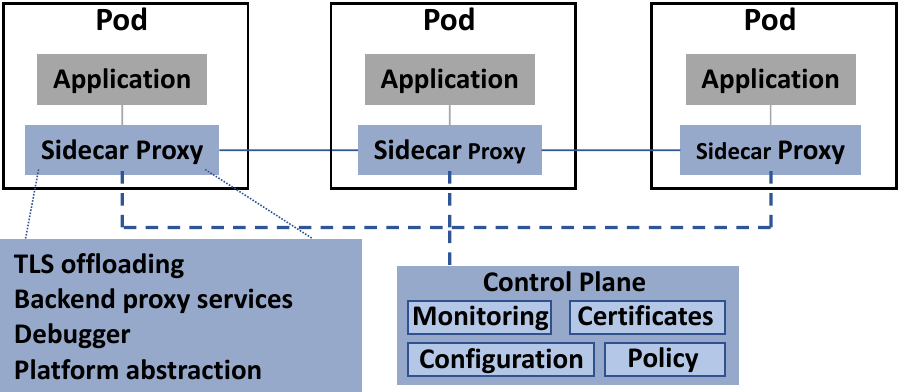}
     \caption{Side-car proxy overview in Kubernetes environment}
     \label{fig:backscp}
\end{figure}

The foundation of our 5G-WAVE integrated platform is built upon the concept of SCPs illustrated in Figure~\ref{fig:backscp}. In a Kubernetes cluster, the smallest logical unit of deployment is a pod. A single pod can be made up of multiple containers that share the same network sandbox.

The SCP is a non-functional container in a pod that provides the main application with a desired set of abstractions for non-functional services such as security procedures, configurations, policies, and monitoring. Combined with a service mesh that connects the SCPs, the deployments can be managed from a centralized control plane. Given their flexible nature, SCPs have become a popular way of augmenting applications with third-party functionality. Furthermore, in 3GPP Release 16, SCPs have been standardized as a method of indirect communication between the 5G core VNFs~\cite{3gpp23501}.

\subsection{Authorization in WAVE}
% WAVE stands for `\textbf{W}ave is an \textbf{A}uthorization \textbf{V}erification \textbf{E}ngine'. 
WAVE is an authorization framework designed for distributed systems to handle transitive delegations and revocations without having a central trust authority~\cite{Andersen2019}. %If an authorization server is compromised, the design of WAVE guarantees that the permissions of uncompromised users are not affected and the adversary cannot see permissions granted in the system, beyond those potentially relevant to the compromised users. 
Initially designed for distributed systems like IoT systems in the built environment \cite{Anderson2018, AndersonThesis}, WAVE can be used as a general-purpose decentralized authorization framework. 
%It allows authorization between \textit{`WAVE entities'} by creating contracts called \textit{`attestations' }for resource access. 
%These attestations are produced during HTTP requests where access is verified before the request is completed. 
%
Based on existing decentralized Trust Management (TM) systems, WAVE improves them with additional security guarantees. The delegations and revocations of permissions are cryptographically enforced, offering confidentiality and invisibility to untrusted parties. 
If an intrusion occurs in the authorization server, the WAVE architecture ensures that the permissions of users who are not compromised remain intact, and the intruder is unable to view any existing system permissions.
WAVE achieves these security guarantees by implementing 1) a Graph-Based Authorization (GBA) model, 2) Reverse Discoverable Encryption (RDE) of attestations, and 3) horizontally scalable untrusted storage.

The WAVE design relies on a GBA model for representing transitive delegations, where the state of the system is maintained in the form of a perspective graph for each entity of the network. Each WAVE \textit{entity} is associated with a unique public-private key pair, representing a user in the system. WAVE achieves decentralized trust by allowing only the transacting entities to grant or revoke permissions through encrypted attestations, which are signed certificates from the \textit{issuer/authorizer} entity to the \textit{subject/requester} entity. Entities form the vertices and attestations form the edges between the authorizer and the requester in the WAVE authorization graph~\cite{andersen2017wave}. An entity which has been granted delegated permissions can form a WAVE proof by traversing the path in the authorization graph from itself to the authorizing entity, discovering the attestations associated with it. Furthermore, any user can verify this WAVE proof, unlike other decentralized trust management systems like SDSI/SPKI~\cite{sdsi1996} and Macaroons~\cite{BirgissonPETVL14}, thus removing reliance on a central verification authority. WAVE objects are stored in horizontally scalable distributed servers to enforce integrity. WAVE has a novel Unequivocable Log Derived Map (ULDM) design for the storage, which scales better than blockchain for addition and retrieval of objects.

% A \textit{namespace} is a domain for a hierarchy of resources that can be accessed by the entities. A root entity is the one which owns the resources and does not require permission from any other entity to access them. The root entity delegates permissions to other entities in its namespace so that they can access chosen resources according to a given policy. In WAVE, delegations are denoted as \textit{attestations}, delivered in the form of encrypted certificates from the \textit{issuer/authorizer} entity to the \textit{subject/requester} entity.
%The entities form the vertices of the WAVE graph and the attestations form the edges between the issuer and subject vertices~\cite{andersen2017wave}. 
%These delegations can either be direct ($A \longrightarrow B$) or transitive ($A \longrightarrow C \longrightarrow B$). When the subject is asked to provide proof of authorization, the proof is obtained by traversing a path of the graph from the subject to the issuer.
%
 %The entities do not interact with a central authority (e.g.,  authorization server in OAuth 2.0) to control access. 
% The attestations are stored in a horizontally scalable untrusted storage that cryptographically enforces confidentiality and integrity. %This storage is based on a novel transparency log called Unequivocable Log Derived Map (ULDM) that allows auditing to verify log integrity. 
%WAVE uses a novel technique called \textit{reverse discoverable encryption} (RDE). This limits log visibility to entities that need them to form proof of authorization. 

%\subsection{WAVE Implementation}
%\textcolor{red}{
Within the scope of our study, we are not concerned with storage mechanisms of WAVE. Our evaluation is focused on the performance of WAVE as a decentralized authorization platform for supporting the 5G core SBA. All operations on the client side are handled by a `WAVE daemon' which acts as an agent to create entities, attestations and proofs, and publishes/retrieves objects to/from persistent storage.
 \label{sec:backg}
%%%%%%%%%%%%%%%%%%%%%%%%%%%%%%%%%
\section{5G-WAVE Integrated Platform} 

% Our goal in designing the 5G-WAVE integrated platform is to demonstrate the integration of decentralized authorization into the 5G core. As a result, the 5G core VNFs can leverage WAVE to enable mutual authorization with other VNFs through permission delegations. To that end, this section will first provide the system overview of the 5G-WAVE integrated platform. Next, a network slice level overview is shown to contextualize the integration for slicing. Following that, an overview of the inter-pod message exchange is shown with sample HTTP transactions. An in-depth description of the intra-pod flow details the roles of the wSCP and rSCP. Finally, the implementation details are discussed. 

%\textcolor{red}{The \tahl{goal of the 5G-WAVE integrated platform is to demonstrate} decentralized authorization integration into the 5G core. This 
The goal of 5G-WAVE is to enable 5G core VNFs to leverage WAVE for mutual authorization via permission delegations. This section provides a detailed overview of the 5G-WAVE system. %along with network slice integration. 
Following that, we describe inter-pod message exchange, intra-pod flow details (wSCP and rSCP roles), and implementation specifics.

\subsection{5G-WAVE Integrated System Overview} \label{sec:5gwavedepov}

\begin{figure}[t]
    \centering
    \includegraphics[width=0.94\columnwidth,trim={11.2cm 6.2cm 8cm 2.8cm},clip]{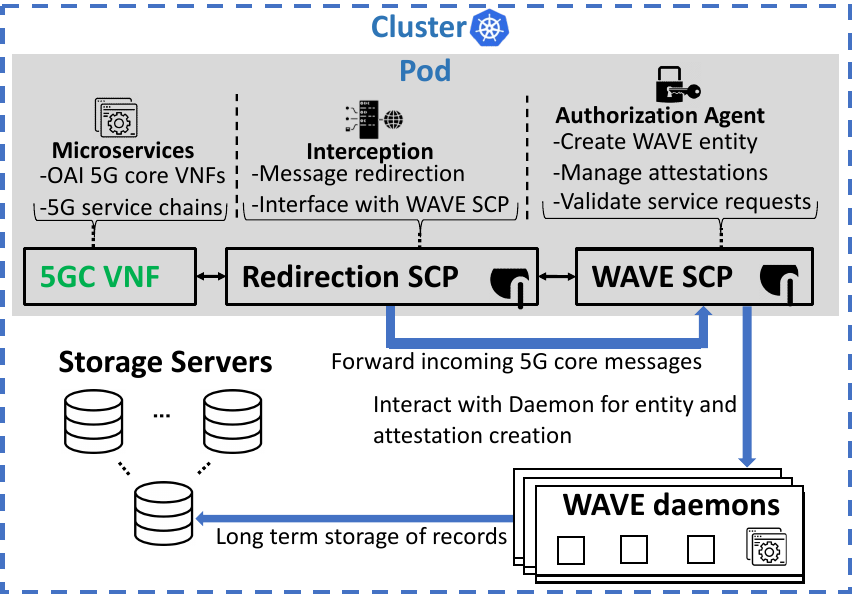}
     \caption{Overview of the 5G-WAVE integrated platform to achieve decentralized inter-VNF authorization in the 5G core}
     \label{fig:wave5gsysov}
\end{figure}

The overview of the 5G-WAVE integration is illustrated in Figure~\ref{fig:wave5gsysov}. Taking place in a Kubernetes cluster, each key component is deployed either at the pod- or cluster-level. To facilitate the interpretation, the components are grouped into three categories: 5G core VNFs, 5G-WAVE integration agents and finally the native WAVE elements. 

As illustrated in Figure~\ref{fig:wave5gsysov}, a single pod is comprised of three different containers. These are the 5G core VNF containers and the 5G-WAVE integration agents (i.e., rSCP and wSCP). The 5G core VNFs are the native OAI software entities~\cite{oaicn5go81online} that are the primary applications of interest in our deployment. To adapt them for this integration, we modified their instantiation process so that they trigger the internal flow of the wSCP (WAVE SCP) before becoming functional. %After the wSCP flow is executed, the 5G core VNFs will resume their standard service chains. %(e.g., user authentication, data session setup, VNF registration). %without requiring additional interactions with the other containers in the same pod.

The rSCP (Redirection SCP) is a custom interception proxy that performs message redirection on incoming HTTP requests. It interacts with the wSCP during the WAVE authorization chain to seek validation on incoming API calls. While the rSCP is a helper to intercept the HTTP messages, wSCP provides functional security features to verify the authorization of the incoming service requests. The wSCP executes the functionalities of a WAVE client on behalf of its respective 5G core VNF. This includes the creation of the WAVE entity as well as the attestations.  %The attestations on the other hand are created on demand when one wSCP (requester) requests authorization from another wSCP (authorizer).

%In this scenario, the two wSCPs are deployed adjacent to each 5G core VNF. The requester wSCP sends the hash of its own WAVE entity along with the authorization request. The authorizer wSCP then utilizes that hash to create an attestation for the requester. Each wSCP grants service access permissions within its own namespace according to the requested policy. For simplicity, we assume the authorization policy to be the same for each wSCP. 

%\textcolor{red}{
The remaining two components are the WAVE daemon and the storage server, both of which are native to WAVE and are deployed at the cluster level. A WAVE daemon serves as an agent responsible for generating entities and attestations and interacts with wSCPs. A WAVE storage server functions as the database where the WAVE daemon publishes entities and attestations. Both WAVE daemons and storage servers can be distributed across multiple logical instances.

%If cross-slice authorization is required, which can arise when network slices share VNFs, this can be handled differently by the WAVE SCPs (D) (How???). Furthermore, such a seamless management of inter-slice communication will simplify security protocols in network slice access control. \textcolor{red}{Finally, all the WAVE SCP interactions will be monitored and controlled through the service mesh control (E).} 

\subsection{5G-WAVE Integrated Network Slicing Overview}

\begin{figure}[t]
    \centering
    \includegraphics[width=0.96\columnwidth,trim={7.5cm 4.9cm 7.3cm 4.5cm},clip]{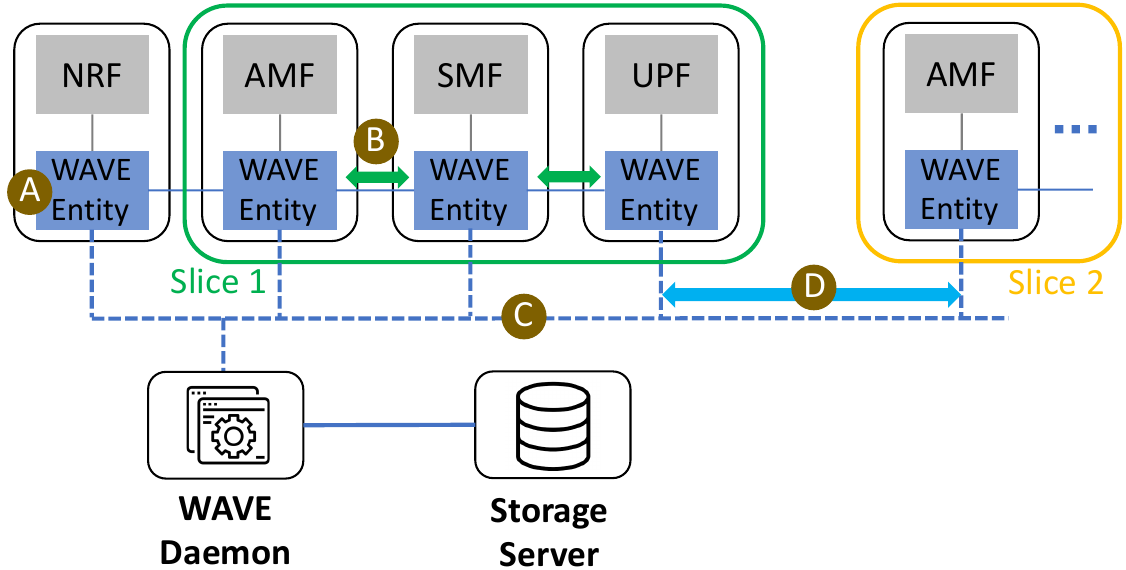}
     \caption{Overview of intra/inter network slice interactions for the integrated 5G-WAVE platform}
     \label{fig:wave5gnetov}
\end{figure}

% With the integration of WAVE into the 5G core, overview of network slice interactions is shown in Figure~\ref{fig:wave5gnetov}. 
An overview of network slice interactions %with WAVE integrated into the 5G core 
is illustrated in Figure~\ref{fig:wave5gnetov}. For simplicity, we have abstracted out the rSCP from the representation of individual pods. Two different network slices are shown with a single NRF and different AMF, SMF, UPFs. In Figure~\ref{fig:wave5gnetov}, (A)-(C) correspond to intra-slice flows while (D) is an inter-slice flow.

For the construction of a network slice in 5G-WAVE, the critical design element is the attachment of a WAVE entity to each 5G core VNF. Thus, for the flows in Figure~\ref{fig:wave5gnetov}, the first step in network slice construction is (A) the creation of the WAVE entity adjacent to each VNF. This occurs inside the wSCP of each pod in Figure~\ref{fig:wave5gsysov}. Hereafter, the VNFs
%5G core applications 
will offload authorization functionality to these WAVE entities residing in each wSCP. During the network slice construction chain, each VNF will (B) authorize the subsequent VNF down the 5G core service chain. For example, for the 5G-AKA procedure, the involved VNFs (i.e., UDR, UDM, AUSF, AMF) will construct an authorization chain using their respective WAVE entities. The WAVE entities belonging to the VNFs of different network slices will be able to (C) communicate with each other using the same WAVE daemon. 

In addition to intra-slice authorizations, (D) inter-slice authorizations can also be granted in the 5G-WAVE integrated platform. This facilitates the communication of VNFs from different network slices without relying on an NRF to distribute access tokens. For network slice handover scenarios of the 5G clients, inter-slice authorization chains can be created to secure the VNF communications. 

\subsection{Authorization Chain Message Exchange}

\begin{figure*}[]
    \centering
    \includegraphics[width=0.8\textwidth,trim={0.2cm 4.3cm 0.7cm 0.6cm},clip]{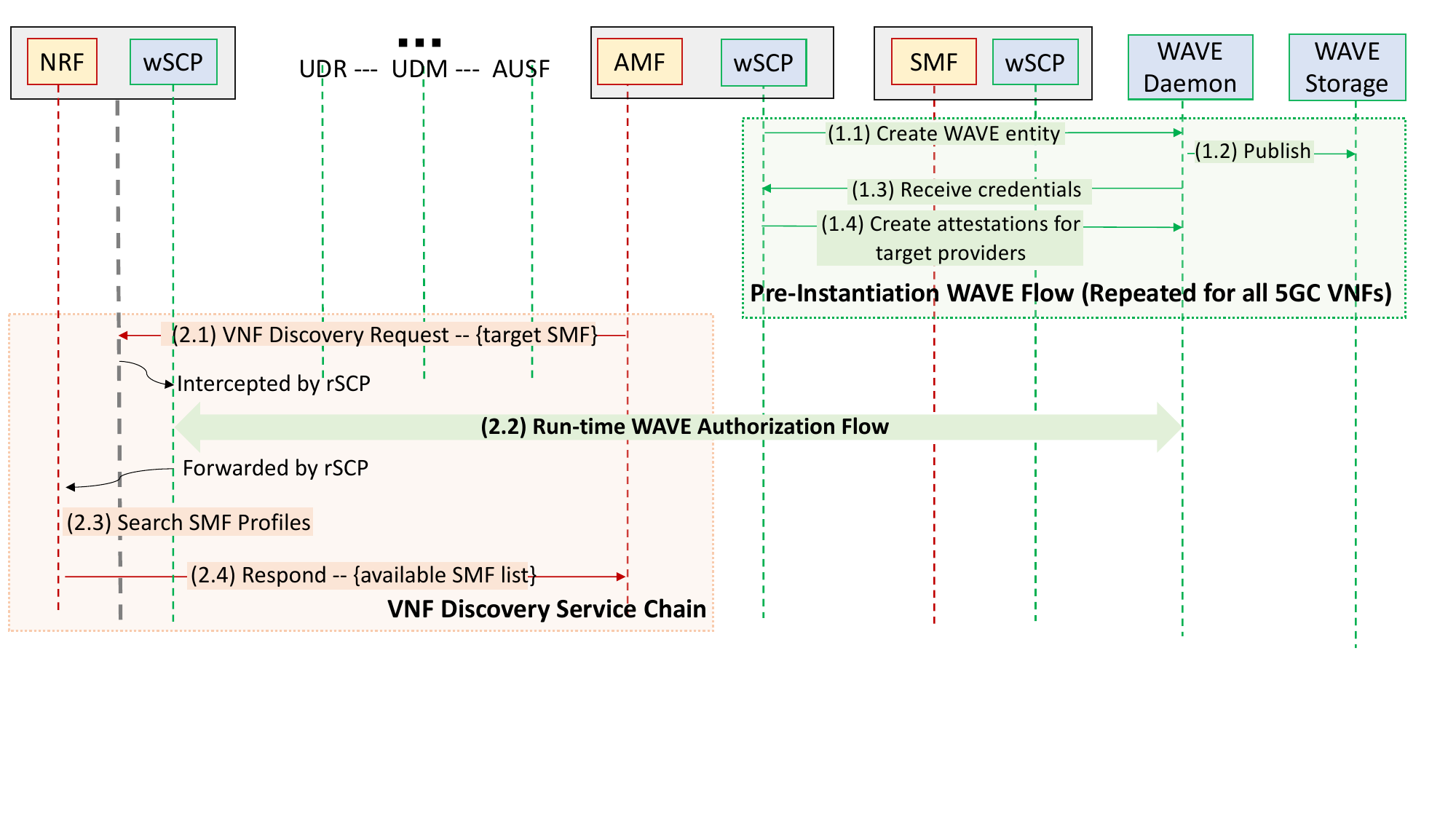}
     \caption{Overview of the new  message flow between WAVE entities and 5G core VNFs for authorization during network slice construction. The rSCP has been abstracted out from the pods for simplicity.}
     \label{fig:wave5gmsgflow}
\end{figure*}

To illustrate how the message flow within the OAI 5G core network is affected, a sample HTTP interaction between the NRF, AMF, and SMF is shown in Figure~\ref{fig:wave5gmsgflow}. For brevity, the rSCP has been abstracted out. Nevertheless, all the incoming messages to a given pod are still intercepted and forwarded through the rSCP. The individual HTTP interactions in this sample scenario have been labeled as belonging to either a 5G core service chain or the internal WAVE authorization flow.

The first set of messages occur during the instantiation of the 5G core VNFs, before they become service-ready. During this pre-instantiation phase, the WAVE flow (1.1)-(1.4) is repeated for all the 5G core VNFs. In Figure~\ref{fig:wave5gmsgflow}, we show it occurring only for the AMF to simplify the illustration. First, (1.1) the WAVE entity is created inside the wSCP and (1.2) published through the WAVE daemon and storage server. The wSCP of the AMF (1.3) receives the necessary WAVE credentials related to its WAVE entity. After the entity creation, it will (1.4) create attestations for service consumers that will be contacting it with service requests.

At this point, the VNF is service-ready and the relevant 5G core service chains can begin. In Figure~\ref{fig:wave5gmsgflow}, we illustrate the VNF discovery service request from AMF. The AMF will (2.1) contact the NRF, seeking a target SMF to set up the network slice. This message is intercepted by the pod sandbox of the NRF so that the authorization can be (2.2) verified through the WAVE flow. Then, the original message is forwarded to the NRF where (2.3) the profile of a candidate SMF is retrieved and (2.4) sent back to the AMF. The same process repeats itself for any subsequent 5G core service chains.

\subsection{wSCP and rSCP Message Flows}

\begin{figure}[b]
    \centering
    \includegraphics[width=\columnwidth,trim={1cm 8.7cm 0.2cm 0.5cm},clip]{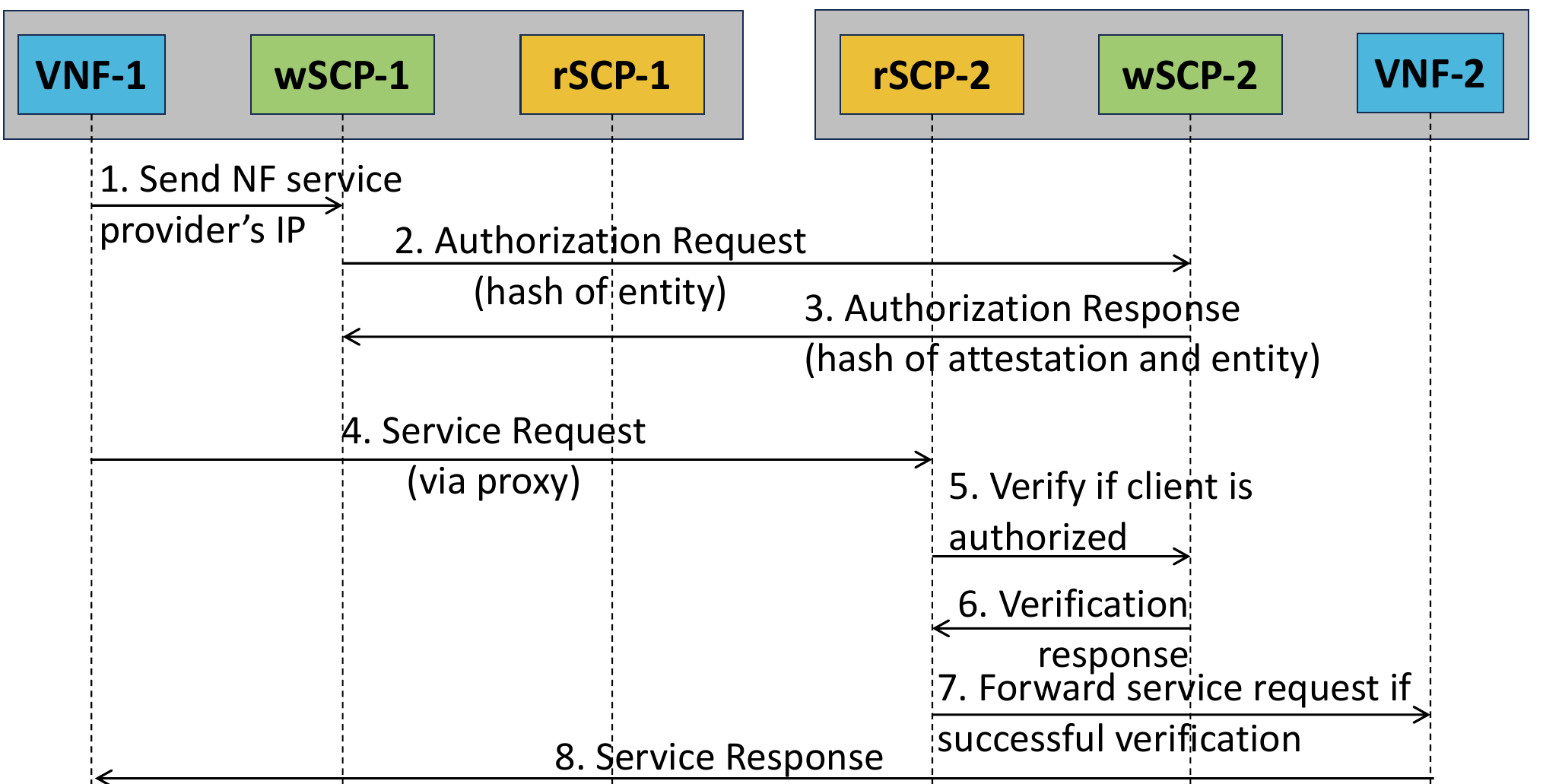}
     \caption{Detailed message flow between the rSCP and wSCP of two 5G core VNFs}
     \label{fig:flowchart}
\end{figure}

The detailed message flow between two sample 5G core VNFs is shown in Figure~\ref{fig:flowchart}. Each pod is made up of the 5G core VNF as well as the wSCP and rSCP. Before the beginning of this message exchange, each wSCP has created its own WAVE entity as described in Section~\ref{sec:5gwavedepov}.

Upon instantiation, \ding{182} VNF-1 will send a list of target service providers to its adjacent wSCP. This will be a list of IP addresses identifying the target VNFs from which VNF-1 will seek to consume a service. For the depiction in Figure~\ref{fig:flowchart}, VNF-1 is seeking this authorization from VNF-2. Hence, after wSCP-1 receives the target list of service providers (i.e., VNF-2), it will contact them \ding{183} with an authorization request. This request will be sent with the hash of the WAVE entity created by wSCP-1. In response, \ding{184} wSCP-2 will send back the hash of the attestation and hash belonging to its own WAVE entity. This acknowledges that VNF-1 has been authorized by VNF-2. 

To initiate the relevant 5G service chain, \ding{185} VNF-1 sends a service consumption request to VNF-2. Upon entering the pod network sandbox, this message is redirected to rSCP-2 instead of directly going to VNF-2. This redirection is achieved by performing port forwarding on the incoming HTTP requests so that they end up at the rSCP, which in turn communicates with the wSCP. Inside rSCP-2, the HTTP transactions are parsed and the source IP address (i.e., the IP of VNF-1) is extracted. The rSCP-2 \ding{186} forwards the IP of the sender to wSCP-2 where it triggers the verification process within the WAVE authorization chain. If the authorization request is successfully verified, the wSCP-2 \ding{187} responds with a valid verification message to rSCP-2 which \ding{188} forwards the original HTTP request from \ding{185} to VNF-2. However, if the authorization response from WAVE is not verified, the service request gets denied and is not forwarded to the 5G component. Once VNF-2 receives the HTTP message, it \ding{189} responds to VNF-1.

%The WAVE SCP offloads the authorization required among the 5GC VNFs. This is achieved by having a WAVE client on each SCP which interacts with the WAVE daemon to create WAVE entities and attestations. The SCPs in turn interact with each other through their individual HTTP servers. As the SCP is created, a WAVE entity corresponding to each gets created too through the WAVE APIs (actionMakeEntity).

%With the framework shown in Figure~\ref{fig:wave5gintow}, the standardized message flow for authorization and discovery shown in Figure~\ref{fig:5gdiscov} no longer requires steps (4)-(6) to be orchestrated by the NRF. Instead, we have the new message flow given in Figure~\ref{fig:wave5gmsgflow}.

% We assume that prior to SMF discovery by the AMF, attestation chain starting from the NRF as the root entity and going through the UDR, UDM and AUSF has already been established. In steps (1)-(2) AMF similarly discovers the SMF. After selecting an SMF (3.1), the authorization task is delegated to the WAVE SCP (3.2). From this point forward, the message flow takes place within the WAVE architecture, starting with the creation of the WAVE entity by the SCP (3.3). The WAVE daemon publishes the entity creation (3.4) and forwards the credentials (3.5) to SCP of AMF. Using these credentials, the WAVE SCP of the AMF will create an attestation (4) for the WAVE entity of the SMF, which will be published in the storage server (4.1). 

\subsection{Implementation Details}
 We use Kubernetes for large-scale deployment of the 5G core VNFs with WAVE entities across multiple Virtual Machines (VMs). The WAVE daemon and storage server are also containerized pods, that have been deployed locally. Even though the storage server can be hosted on the cloud, we opt for a local deployment to reduce the latency of communication between the storage server and WAVE daemon. 

 \textbf{5G core VNFs.} We use v1.2.1 of the OAI 5G core~\cite{OpenAirI28online} VNFs for implementation. We also introduce an additional code block into each VNF's instantiation to trigger the authorization request for attestation from target providers wSCP.  

\textbf{Design of wSCPs.} We design the wSCPs from scratch as independent HTTP servers built using C++17. %where we use the Pistache~\cite{pistache72online} library for creating the HTTP server. 
In addition to the HTTP server, the wSCP container is also populated with the relevant WAVE scripts for the creation of WAVE entities and attestation. We use WAVE binaries \textit{waved} and \textit{wv} from the WAVE open-source repository \cite{immesys2022Oct} as WAVE daemon and agent to create WAVE objects in each wSCP. The scripts are executed at the relevant endpoints of the wSCP HTTP server. 

\textbf{Design of rSCPs.} The rSCP is another HTTP server. However, to facilitate the manipulation of the HTTP messages, we use Python to implement the rSCP. We chose to separate the rSCP and wSCP in this manner to further decentralize the deployment based on microservice design principles. This allows for a more flexible production-grade deployment where the different sub-processes of the 5G-WAVE integrated platform are executed by independent entities. 

\textbf{Pod sandbox initContainer.} %The network sandbox of the pods are modified upon instantiation using an initContainer. 
We modify the IPTABLES of the pod using an initContainer so that %any 
incoming/outgoing messages to/from the %5G core 
VNFs are redirected to the rSCPs.

 \label{sec:framework}
%%%%%%%%%%%%%%%%%%%%%%%%%%%%%%%%%

\section{Experimental Setup} \label{sec:expsetup}
\begin{figure}
    \centering
    \includegraphics[width=0.8\columnwidth,trim={9.3cm 6.6cm 12.5cm, 4cm},clip]{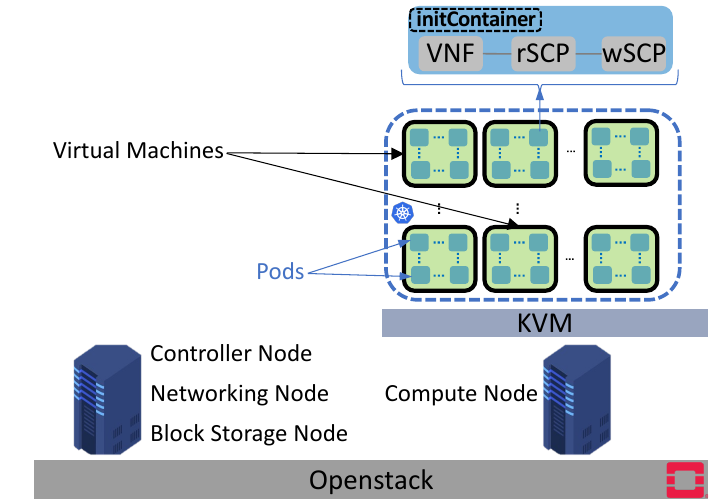}
    \caption{Testbed with Openstack virtual machines hosting a Kubernetes cluster}
    \label{fig:expsetup}
\end{figure}

Our experimental setup is shown in Figure \ref{fig:expsetup}. 
The infrastructure is comprised of two physical nodes that are hosting an OpenStack orchestrator. One of the nodes is the controller, networking and block storage node while the other serves as a compute node with the KVM hypervisor. The physical nodes used are two Precision 7920 Tower servers with
%\begin{itemize}
  %  \item 
  2 x Intel Xeon Gold 5218R 2.1GHz CPUs,
    %\item 
    512GB RAM,
    %\item 
    1TB disk space, and running
    %\item 
    Ubuntu 20.04.
%\end{itemize}

The compute node is running a total of 14 VMs, which are hosting a Highly Available (HA) Kubernetes cluster for production-grade testing. The roles and the flavors of the VMs in the cluster are given in Table~\ref{tbl:kubclusconf}.

\begin{table}[b]
% \vspace{12pt}
\centering
\small\selectfont
\caption{HA Kubernetes cluster VM flavors}
\label{tbl:kubclusconf}
%\renewcommand{\arraystretch}{1.1} % for the vertical padding
%\resizebox{\textwidth}{!}{
\begin{tabular}{c|c|c|c|c}
%\hline
\textbf{Node} & \textbf{Instances} & \textbf{vCPUs} & \textbf{RAM (GB)} & \textbf{Disk (GB)} \\\hline \hline
\textbf{Control} & 2 & 2 & 8 & 40 \\%\hline
\textbf{Worker} & 11 & 4 & 8 & 80 \\%\hline
%\textbf{l-workers} & 2 & 20 & 8 & 80 \\%\hline
\textbf{Storage} & 1 & 4 & 8 & 80 \\\hline
%L-worker1  & 1 & 20 & 200 & 80 \\\hline
%L-worker2  & 1 & 15 & 100 & 80 \\\hline
\end{tabular}
%}
\end{table}

Our 5G core deployment consists of 6 control plane VNFs (i.e. NRF, UDR, UDM, AUSF, AMF, SMF) and 1 user plane VNF (i.e., UPF) in each slice deployed as Kubernetes pods. Each pods consists of 3 containers: 5G core VNF, wSCP and rSCP. We use the v1.2.1 OAI implementation of the 5G core with gNBSIM~\cite{RohanGnb17online} entity for the RAN. 
%We integrate the WAVE side-car proxies only on pods corresponding to these VNFs. 
 %We deploy the 5G core framework for two separate settings: one with wSCP integrated into the VNF pods for authorization, and the other without. 
 %Currently, the OAI 5G core (or any other open-source 5G implementation) has no native authorization functionality built for it. %Each experiment is run for 20 iterations.

  \label{sec:experiment}
%%%%%%%%%%%%%%%%%%%%%%%%%%%%%%%%%

\section{Performance Evaluation}

\begin{figure}[t!]
 %\vspace{-0.in}
    \centering
    \includegraphics[width=0.8\columnwidth, trim={5cm 5cm 12cm 0.2cm},clip]{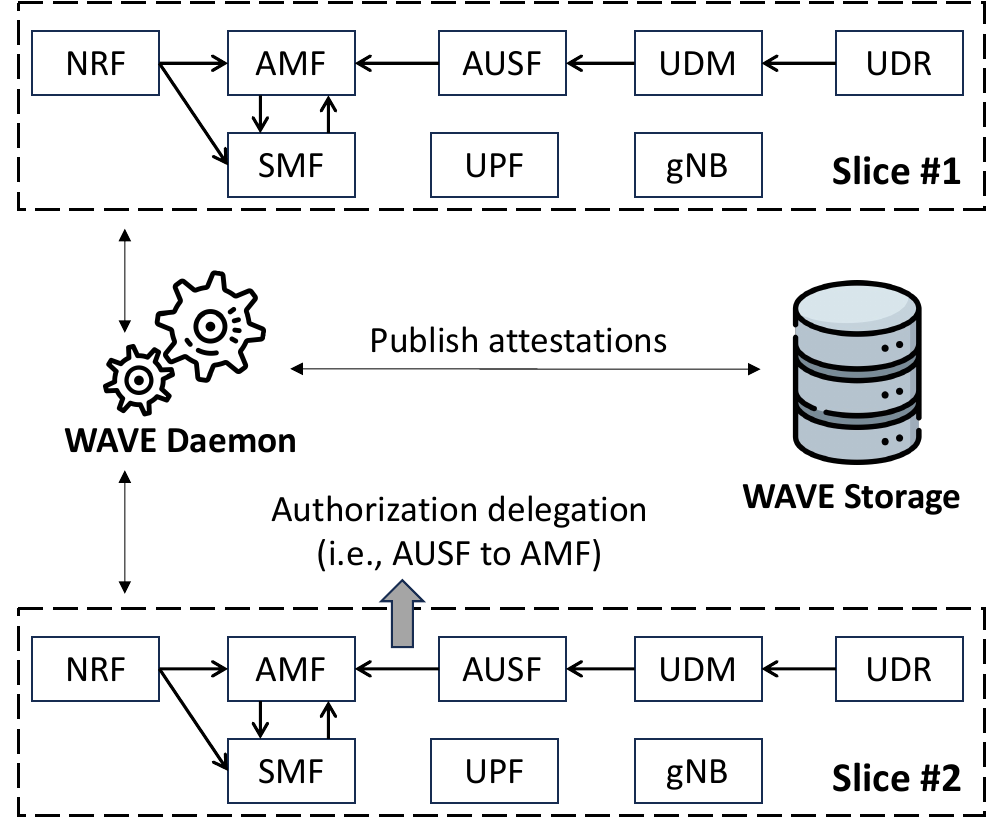}
     \caption{Authorization chains in two different network slices}
     \label{fig:slicetopology}
\end{figure}

%\textcolor{red}{
We conduct two sets of experiments to evaluate the performance of 5G-WAVE. In particular, we measure the HTTP request-response times from two perspectives: 1) authorization overhead in single slice and 2) scalability in multiple slice deployment.
For the first set of experiments, we deploy a single slice of the 6 core VNFs in the control plane along with a UPF and a gNBSIM. %Our goal is to measure the authorization overhead of the 5G-WAVE integrated platform compared with the vanilla 5G deployment.
For the second set of experiments, we increase the number of network slices to measure the authorization overhead with more VNFs and analyze how 5G-WAVE scales with network size. Within each network slice, there is a VNF authorization chain. For scalability tests, we deploy multiple logically-isolated slices sequentially for use cases with strict security requirements. Two such slices are illustrated in Figure \ref{fig:slicetopology}. We perform 20 and 10 iterations, respectively, for the two experiments to average our results.

 %To make the framework parallel to OAuth 2.0, we use the single delegation structure in WAVE, however it is important to note that WAVE provides an additional advantage of transitive delegation over OAuth. 

\begin{table}[b!]
    \centering
    \small\selectfont
    \caption{Time taken (in ms) by authorizer VNFs to create attestations for requesting client VNFs}
    \begin{tabularx}{0.7\columnwidth} {C{0.15\columnwidth}|C{0.15\columnwidth}|Z|Z}
        \textbf{Requester (Client)} & \textbf{Authorizer (Host)} & \textbf{Mean} & \textbf{Std} \\ \hline \hline
         AMF & NRF & 156.52 & 12.35 \\
         SMF & NRF & 133.16 & 17.54 \\
         UDM & UDR & 159.40 & 28.64 \\
         AUSF & UDM & 197.80 & 21.72\\
         AMF & AUSF & 184.90 & 18.26\\
         SMF & AMF & 195.90 & 21.98\\
         AMF & SMF & 183.94 & 15.94\\
         \hline
    \end{tabularx}
    \label{tab:att_times}
\end{table}

\begin{figure*}[t!]
\vspace{-0.1in}
    \centering
        \includegraphics[width=0.85\textwidth]{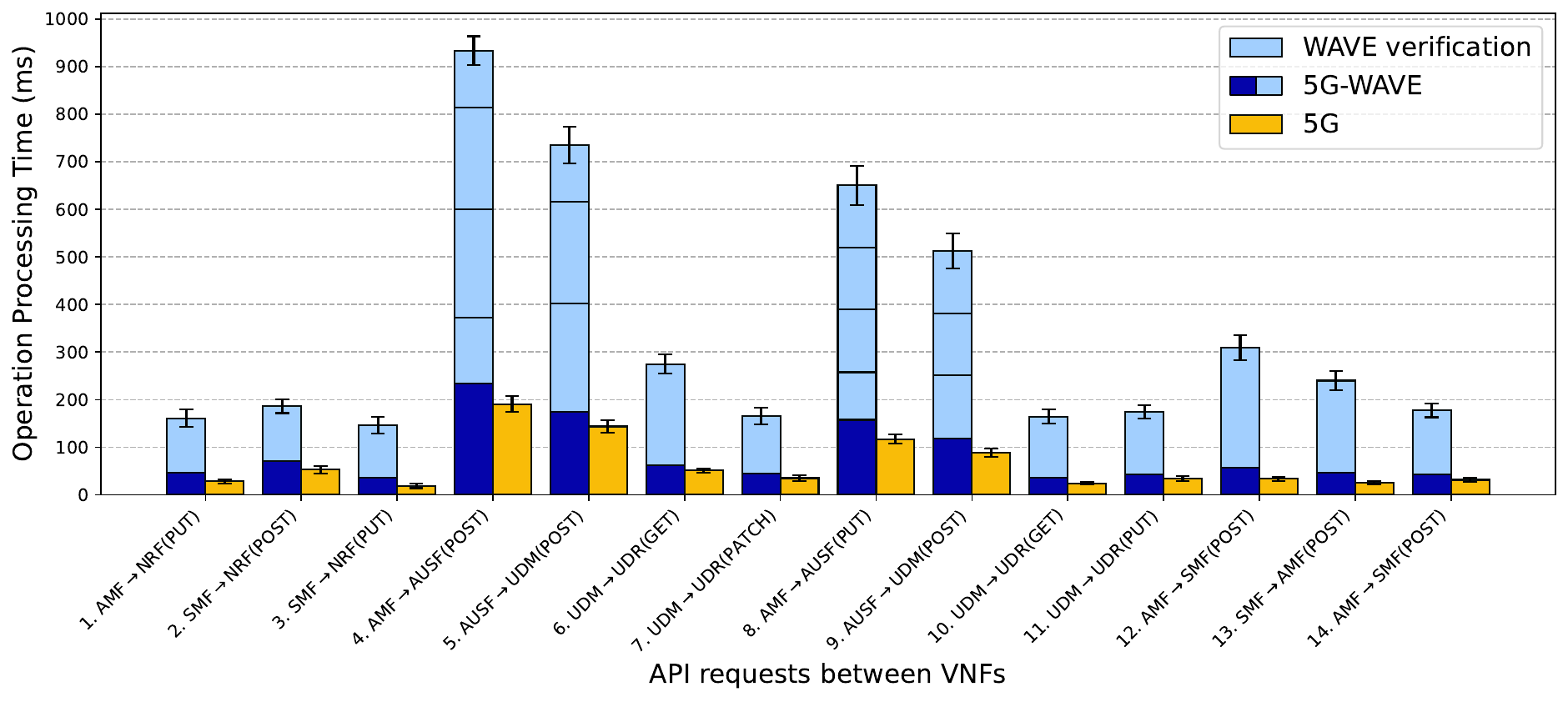}
     \caption{Latency overhead comparison of the 5G-WAVE integrated platform with the OAI 5G core deployment.}
    \label{fig:latency}

\end{figure*}

\subsection{Single slice authorization overhead}
%As of the time of writing this paper, 
So far, none of the open-source 5G core implementations have OAuth 2.0 authorization functionality built into them. Thus, the performance overhead of 5G-WAVE is compared against a 5G deployment with no built-in authorization. 

While analyzing the message flows between VNFs, we identify 7 unique pairs of requester-authorizer VNFs for attestation creation. The mean and standard deviation of the time taken to create attestations for each pair are given in Table \ref{tab:att_times}. The client VNF forms the requesting entity in WAVE and the host VNF forms the authorizing entity. We further identify 14 distinct API requests for a functional deployment of the 5G core, which are listed in Table~\ref{tab:latency}. Each request involves a WAVE verification operation by the authorizing entity. We present the duration of a single verification operation in Table~\ref{tab:latency} and the total overhead for the HTTP request in Figure~\ref{fig:latency}. %The processing time of these requests is compared against time taken in vanilla 5G deployment.
%\textbf{

In Figure~\ref{fig:latency}, two sets of measurements are stacked for each bar representing 5G-WAVE. The results show the time cost for 1) the `verification' operation of WAVE (stacked on top) and 2) 5G core service chain and rerouting overhead via rSCP (stacked on bottom). The majority of the overhead is introduced from the WAVE verification flows. The transactions in the 5G core can be categorized into three groups: 1) VNF instance registration (bars 1-3), 2) 5G-AKA service chain (bars 4-11), and 3) Packet Data Unit (PDU) session setup (bars 12-14). The WAVE verification bars are split further for the 5G-AKA operations to illustrate the number of verification operations in each API request and the time cost for them.

\begin{table}[b!]
\centering
% \vspace{5mm}
\caption{Time taken (ms) for verification of WAVE authorization for different Client$\rightarrow$Host pairs, measured on the authorizing entity's end. The Client is the requesting WAVE entity and the Host is the authorizing WAVE entity.}
\label{tab:latency}

\begin{tabularx}{\columnwidth}{|m{0.02\columnwidth}|m{0.07\columnwidth}|m{0.07\columnwidth}|Z|m{0.08\columnwidth}|}
    \hline
     & \textbf{Client} & \textbf{Host} & \textbf{Service Request} & \textbf{Time}  \\
    \hline
    \multicolumn{5}{|l|}{\textit{VNF instance registration}} \\
    \cline{1-5} 
    \hline
    1 & AMF & NRF & PUT: nnrf-nfm (NF registration) & 114.55 \\
    \hline
    2 & SMF & NRF & POST: nnrf-nfm (Subscribe to updates)& 115.65 \\
    \hline
    3 & SMF & NRF & PUT: nnrf-nfm (NF registration) & 110.75 \\
    \hline
    \multicolumn{5}{|l|}{\textit{Authentication and Key Agreement}} \\
    \cline{1-5} 
    4 & AMF & AUSF & POST: nausf-auth (UE authentication) & 138.80 \\
    \hline
    5 & AUSF & UDM & POST: nudm-ueau (Generate authentication vectors) & 227.95 \\
    \hline
    6 & UDM & UDR & GET: nudr-dr (Query UE authentication data) & 212.65 \\
    \hline
    7 & UDM & UDR & PATCH: nudr-dr (Update UE authentication data) & 120.25 \\
    \hline
    8 & AMF & AUSF & PUT: nausf-auth (Authentication confirmation) & 99.90 \\
    \hline
    9 & AUSF & UDM & POST: nudm-ueau (Inform about authentication result) & 132.30 \\
    \hline
    10 & UDM & UDR & GET: nudr-dr (Query UE authentication data) & 129.05 \\
    \hline
    11 & UDM & UDR & PUT: nudr-dr (Update UE authentication data) & 131.60 \\
    \hline
    \multicolumn{5}{|l|}{\textit{PDU Session Setup}} \\
    \cline{1-5} 
    12 & AMF & SMF & POST: nsmf-pdusession (Create SM Context) & 252.00 \\
    \hline
    13 & SMF & AMF & POST: namf-comm (N1-N2 message transfer) & 194.10 \\
    \hline
    14 & AMF & SMF & POST: nsmf-pdusession (Update SM Context) & 135.55 \\
    \hline
    \multicolumn{4}{|r|}{\textbf{Mean Time}} & 151.08 \\
    \cline{1-5} 

\end{tabularx}
% \vspace{-1mm}

\end{table}

\textbf{VNF registration.} During instance registration, both AMF and SMF send requests to NRF. We see from Table~\ref{tab:latency}, that the WAVE verification process takes almost the same amount of time for the three HTTP requests ($\approx$ 113 ms). Looking at Figure~\ref{fig:latency}, these correspond to the PUT requests going from the AMF/SMF to the NRF and incur the lowest overhead among all the HTTP transactions.

\textbf{5G-AKA service chain.} This is the primary UE authentication service chain that comprises a series of API requests between the AMF, AUSF, UDM, and UDR. This service chain is triggered when the UE sends a registration request to the AMF. The AMF interacts with the AUSF which in turn requests the UDM to generate the authentication vector, which further propagates it to the UDR. This is a service chain of three VNFs that involves 4 sequential API transactions for the original UE authentication request to be completed. Thus, we can see in Figure~\ref{fig:latency}, in the \texttt{AMF$\rightarrow$AUSF(POST)} and \texttt{AMF$\rightarrow$AUSF(PUT)} operations, the WAVE verification duration is the sum of four verification operations. Although the UE authentication has \texttt{4x} verification overhead,  it is a one-time process and the security benefit of WAVE outweighs the overhead in total request completion time.

\textbf{PDU session setup.} %Orchestrated by the AMF and SMF, %the 
PDU session setup consists of two \texttt{AMF$\rightarrow$SMF(POST)} API calls. Compared to the 5G-AKA service chain, the WAVE overhead is smaller for PDU session setup. A single UE can instantiate multiple PDU sessions within the same network slice. This overhead will not introduce significant delays in the 5G core operations.

\subsection{Scalability overhead with multiple slices}
\begin{figure}[t!]
    \centering
    \vspace{-3mm}
    \includegraphics[width=0.96\columnwidth, trim={0cm 0cm 0cm 0cm},clip]{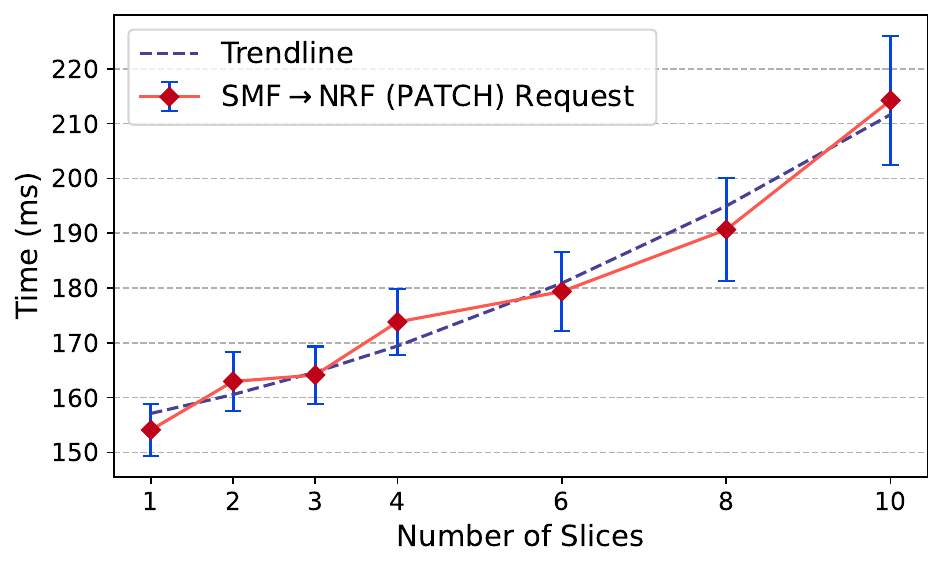}
     \caption{Latency overhead with increasing number of slices}
     %\vspace{3mm}
     \label{fig:scale-measure}
\end{figure}

To evaluate how the 5G-WAVE framework scales in a larger network, we deploy multiple slices which are isolated from each other. After VNF instance registration, SMF sends a heart-beat timer request to NRF periodically. This is a \texttt{PATCH} HTTP request that involves a single WAVE verification operation. We measure the average time taken for this request to complete as we keep increasing the number of slices in the deployment. This gives us the estimate of how long 5G-WAVE will take for a single HTTP message. It can be observed from Figure~\ref{fig:scale-measure} that for a single slice deployment, the overhead by 5G-WAVE is 155ms. We further observe a linear trend in the increase of request completion time with the number of slices. This is expected because there is an increase in computational demand from the WAVE daemon with more slices. Note that even when the number of slices grows by 10x, the time taken only increases by 1.4x. Thus, we can claim that 5G-WAVE scales well with the network size. 

 \label{sec:perfeval}
%%%%%%%%%%%%%%%%%%%%%%%%%%%%%%%%%

\section{Security Analysis}

In this section, we describe our two-stage security analysis of the 5G-WAVE integrated platform. Firstly, we discuss how specific OAuth attack vectors are addressed in Table~\ref{tbl:oauthvswwave}. Secondly, to illustrate how the proposed design improves the flexibility and security of the 5G core, we reference Key Issues (KIs) from the 3GPP standardization in Table~\ref{tbl:keyissues}. 

\subsection{Comparing 5G-OAuth and 5G-WAVE}

\begin{table}[b]
\centering
% \small\selectfont
\caption{Comparing the detection and mitigation capabilities of the standardized 5G-OAuth framework with the 5G-WAVE integrated platform ($\checkmark$: addressed, \FilledTriangleUp: addressed with additional client-side configuration, x: vulnerable)}
\label{tbl:oauthvswwave}
\begin{tabular}
{l|c|c}
%\hline
\textbf{Attack Type} & \textbf{5G-OAuth} & \textbf{5G-WAVE} \\ \hline \hline
%\multicolumn{1}{|c|}{}                                 & \multicolumn{1}{c}{\scriptsize \textbf{Detect}} & \multicolumn{1}{c|}{\scriptsize \textbf{Mitigate}} & \multicolumn{1}{c}{\scriptsize \textbf{Detect}} & \multicolumn{1}{c|}{\scriptsize \textbf{Mitigate}} \\ \hline    
\small Authorization Code Injection & \cellcolor{orange!40}\FilledTriangleUp & $\cellcolor{green!10}\checkmark$  \\ %\hline
\small Access Token Leakage & \cellcolor{orange!40}\FilledTriangleUp & \cellcolor{green!10}$\checkmark$ \\ %\hline    
\small Credential Phishing Attacks & \cellcolor{orange!40}\FilledTriangleUp  & \cellcolor{green!10}$\checkmark$ \\ %\hline
\small Authorization Flooding  & \cellcolor{red!30}x & \cellcolor{green!10}$\checkmark$  \\ %\hline
\small Access Token Hijacking & \cellcolor{orange!40}\FilledTriangleUp  & \cellcolor{green!10}$\checkmark$  \\ \hline
%\small Service Scanning & \cellcolor{green!10} $\checkmark$ & \cellcolor{green!10} $\checkmark$  \\ \hline
\end{tabular}
\end{table}

To ensure inter-slice and intra-slice security, 5G core SBA must adopt a Zero-Trust Architecture (ZTA) \cite{Rose2020Aug}, which demands all network entities to be authenticated, authorized, and continuously verified before being granted access to resources. 3GPP has standardized OAuth 2.0 as an authorization framework for VNF service access to achieve SBA domain security. Within this framework, the NRF acts as an authorization server that distributes access tokens to service consumer VNFs~\cite{3gpp33501,3gpp29510}. Either directly or as a result of misconfiguration, OAuth 2.0 is vulnerable to the attack types listed in Table~\ref{tbl:oauthvswwave}.

When OAuth 2.0 is replaced with a decentralized approach, we remove the dependency on the NRF as a central authorization server. Each VNF service provider becomes both the authorization and resource server for their own services. To realize this idea, our decentralized authorization framework of choice is WAVE. It has been shown that WAVE performs better than traditional authorization frameworks by providing stronger security guarantees with comparable performance~\cite{Andersen2019}. %It fulfills the 5G core security requirements \cite{3gpp33501} while increasing the flexibility of deployment. 

In WAVE, services are granted in the form of encrypted attestations rather than tokens. As a result, the resource requester and resource provider are bound to each other in the WAVE authorization chain without involving a third-party server. These attestations can be stored in a distributed manner using horizontally scalable servers without compromising system security. The critical security features provided by incorporating WAVE into the 5G core are listed below.

\begin{itemize}
    \item Confidentiality, integrity, and replay protection are supported for inter-VNF communication.
    \item The 5G core network slice topology in each administrative domain is maintained separately. Thus, for Multi-Operator Networks (MNOs), the VNF topologies are hidden from entities in different trust domains.
%    \item Service consumers/providers are mutually authenticated. 
    \item Instead of relying on an authorization server, the service request messages will be validated by the service provider VNFs. Invalid or unauthorized messages will be rejected or discarded according to the protocol specification.
\end{itemize}

A comparison of the 5G-OAuth and 5G-WAVE is provided in Table~\ref{tbl:oauthvswwave}. The specific attack vectors are explained below.

As a Man-in-the-Middle (MiTM) attacker, adversaries can perform authorization code injections, %\textbf{
access token hijacking and take advantage of %\textbf{
token leakages in the OAuth 2.0 framework~\cite{RFC6819O63online,draftiet79online,yang2016model}. Without additional client-side security implemented on the VNFs (e.g., URL redirects, TLS), the 5G core remains vulnerable to these OAuth 2.0 attack vectors. On the other hand, WAVE prevents these attacks by removing the reliance on access tokens. By extension, this no longer requires the employment of risk-inducing information elements such as authorization codes and access scopes.

As a centralized authorization server, the NRF is vulnerable to authorization flooding attacks. Adversaries can execute a Distributed Denial of Service (DDoS) by generating a high volume of access token requests directed at the NRF. With the decentralization provided by the 5G-WAVE integration, this vulnerability (i.e., NRF as OAuth 2.0 server) is removed. 

Adversaries using phishing attacks will impersonate a legitimate service requester to obtain illegal access tokens and/or access credentials from legitimate OAuth 2.0 entities. However, WAVE uses custom entities adjacent to each 5G core VNF to ensure their authenticity. As a result, any impersonator not accompanied by a WAVE entity will not be able to infiltrate the authorization chain of the 5G-WAVE integrated platform.

Several attacks from Table~\ref{tbl:oauthvswwave} can be addressed by additional client-side configuration. However, this requires additional policies, development effort, and expertise. With the 5G-WAVE integrated platform, all the attack types are addressed out of the box. This enables mobile operators to focus on deployment without concern about the underlying security.

\begin{table}[t]
\centering
% \small\selectfont
% \vspace{5mm}
\caption{Summary of 3GPP key issues addressed by the 5G-WAVE integrated platform ($\checkmark$: directly addressed, \FilledTriangleUp: addressed with proper configuration, $\hourglass$: future work)}
\label{tbl:keyissues}

% \begin{tabular}{p{0.1\columnwidth}|p{0.3\columnwidth}|p{0.2\columnwidth}}
\begin{tabular}{c|c|c|c}
 \textbf{3GPP \#} & \textbf{KI \#} & \textbf{Description} & \textbf{5G-WAVE} \\
 \hline \hline
 33.813~\cite{3gpp33813} & 5 & token handling between slices & \cellcolor{green!10} $\checkmark$ \\
 33.886~\cite{3gpp33886} & 2 & temporary slice authorization & \cellcolor{orange!40} \FilledTriangleUp \\
 33.874~\cite{3gpp33874} & 3 & AF authorization & \cellcolor{orange!40} \FilledTriangleUp \\
 33.813~\cite{3gpp33813} & 1 & slice-specific authorization & \cellcolor{red!30} $\hourglass$ \\
 \hline
\end{tabular}
% \vspace{-1mm}
\end{table}

\subsection{Addressing 3GPP Key Issues}

\textbf{[33.813 KI 5]} The NRF can be deployed at different administrative hierarchies (e.g., Public Land Mobile Network level, shared-slice level, slice-specific level). The tokens distributed by the NRF can be used to access services of the same type of VNF service providers across multiple slices. This can lead to illegal access requests since different network slices may have alternative service consumption policies for their VNF types. The 5G-WAVE integrated platform directly addresses this KI as it removes the need for tokens altogether.

\textbf{[33.886 KI 2]} For the graceful termination of data sessions, UEs are allowed access to temporary slices. This requires efficient authorization mechanisms to moderate access to such slices. By creating a joint authorization chain between the UE's current and target temporary slice, authorized access can be provided with the 5G-WAVE integrated platform. 

\textbf{[33.874 KI 3]} External Application Functions (AFs)~\cite{3gpp29517} can be exposed to the 5G core SBA for carrying out custom operator functions. Without proper authentication and authorization, malicious AFs can gain access to sensitive information by accessing network slices. The operator can attach the AFs to the 5G-WAVE integrated authorization chain ensuring that the service provider VNFs can directly provide authorization to AFs for service consumption.

\textbf{[33.813 KI 1]} Network slice-specific authentication and authorization has become a fundamental requirement. To address this KI, 3GPP has already standardized the Network Slice-Specific Authentication and Authorization Function (NSSAAF)~\cite{3gpp29526}. This additional VNF will serve as an independent authentication and authorization server for moderating access to individual slices. Currently, the 5G-WAVE integrated platform cannot directly be used to address this KI. As a future work, we plan to store slice-specific metadata inside the wSCPs to keep track of the VNF authorizations along with their network slices. 

% \vskip\baselineskip

 \label{sec:seceval}

 \label{sec:discussion}
%%%%%%%%%%%%%%%%%%%%%%%%%%%%%%%%%

\section{Related Work}

With the growing popularity of open-source 5G, various testbed-based studies have spawned investigating several scalability aspects~\cite{atalay2022network,atalay2022scaling}. Furthermore, as the microservice model of deployment for applications has gained more traction, authorization methodologies have adapted to accommodate the growing volume of interactions. 

% \subsection{5G Authorization}
%For the new 5G service-based architecture~\cite{3gpp23501}, 3GPP has standardized the NRF~\cite{3gpp29510}. The NRF orchestrates interactions between VNFs by acting as a local repository for them to register with and later discover other VNFs in the service chain. Furthermore, it distributes access tokens, enabling them to securely exchange information elements over HTTP transactions. Several novelties have been proposed to enrich the functionalities of this base model of enabling authorization.

% The majority of the existing 5G studies have focused on enhancing the performance of specific control or user plane procedures. Prominent studies~\cite{xu2022tutti,kim2022outran,moradi2018softbox} have focused on improving the 5G core performance to achieve low latency and reducing inter-VNF communication overhead rather than augmenting the security of 5G core.

% A low latency 5G edge video analytics framework is proposed in~\cite{xu2022tutti}. To reduce inter-VNF communication overhead, ~\cite{jain2022l25gc} has proposed a low-latency 5G control plane design that bypasses REST APIs through shared memory communication between VNFs. Similar prominent studies~\cite{kim2022outran,moradi2018softbox} have all focused on improving the 5G core performance rather than augmenting its security.

Focusing on the security aspects of 5G, Edris et al.~\cite{edris2020network} proposed a single sign-on federated identity for allowing multiple users to access services offered by providers. Similarly, a higher level authorization is proposed in~\cite{wong2017virtualized}, which focuses on handling authorization for a multi-tenant and operator ecosystem with a distributed and virtual authorization agent. These approaches do not investigate the authorization between individual VNFs but rather focus on the authorization of users and service providers at a higher level. 

To defend against the reusability of access tokens, Zhang et al.~\cite{zhang2021setoken} proposed to use a trusted NRF as a certificate authority to further vouch for the access tokens with a certificate. While the approach adds another layer of security, it further centralizes the authorization mechanism of the 5G architecture. Furthermore, token hijacking and interception attack vectors, as well as authorization flooding attacks, are not addressed. 

A different approach is presented in~\cite{behrad20195g} for network slice specific authorization. Access control for specific slices is offloaded to the 3\ts{rd} party UE operators, instead of relying on the 5G core network. Their work is orthogonal to ours as they focus on UE authentication and access control (AAC) and design additional RAN NFs to delegate control to the 3\ts{rd} parties. By contrast, we address mutual authorization between VNFs for service consumption.
% Even though it decentralizes the authorization servers for certain slices, this approach still suffers from relying on a single third-party authorization server for such a task. Additionally, it considers authorization in terms of network slices, instead of individual VNFs within a slice. %Blockchain is also proposed for authorization in 5G and beyond \cite{blockchain5g}.

%\textcolor{red}{Authors of
Akon et al.~\cite{Akon5Gcore} conducted a formal analysis of OAuth 2.0-based access control mechanism in the 5G core as outlined in the 3GPP specifications, and they identified 5 major vulnerabilities in the design. This work further strengthens our motivation to remove a central authorization server in the 5G core. A very brief discussion of distributed authorization in the 5G micro-service architecture is offered in~\cite{guija2018}, but it lacks implementation and evaluation details.

Looking into 5G and beyond, Atalay et al.~\cite{atalay2022securing} proposed an authorization mechanism for the the O-RAN ecosystem~\cite{ORANALLI88online}. Their work similarly relies on SCPs for offloading functional security from the main applications. Therefore, our proposed 5G-WAVE integrated platform can be easily applied to their design for enhanced security features.

% \subsection{Decentralized Authorization and WAVE}
% Spawnpoint \cite{spawnpoint} is a container orchestration system which combines Docker and WAVE to deploy, run, and monitor microservices. Spawnpoint uses Bosswave~\cite{andersen2017wave}, which leverages blockchain for storage, to provide security guarantees to users.  By contrast, our proposed extension relies on a Unequivocable Log Derived Map (ULDM) based storage of WAVE v3.

%Moreover, Spawnpoint is evaluated by deploying microservices in a built environment, whereas, our work focuses on services in 5G NFV with multiple slices.

% DeFIRED \cite{Vrielynck2022Apr} is a decentralized framework built on top of WAVE that allows users to acknowledge and deny delegations received from the issuing entity. %It also allows the client to publish a revocation for themselves, something which can be done only by the issuer in WAVE. They introduce invitations which are delegations issued by the issuer and attestations represent receiver's acknowledgement of the invitation from the issuer. 
% They benchmark their results against WAVE and show that their method only introduces a small overhead compared to WAVE. %Our work is orthogonal to this and WAVE can be replaced with DeFIRED in our setup since the overhead in both is comparable. 

To the best of our knowledge, no existing work explores application of entity-based decentralized authorization in the 5G network architecture. Our work is the first to present the integration of decentralized authorization into the 5G core. %and measure the overhead of employing decentralized authorization in network slice construction.
 \label{sec:relwork}
%%%%%%%%%%%%%%%%%%%%%%%%%%%%%%%%%

\section{Conclusion}

5G networks have already been deployed globally, and a continued reliance on a central authorization server in 5G core will lead to security and performance issues. To address this, we introduced the 5G-WAVE integrated platform for decentralizing inter-VNF communications. %in the 5G core SBA. 
Our design utilizes an interception/authorization pipeline based on SCPs attached to individual 5G core VNFs deployed as Kubernetes pods. We evaluated the overhead of our proposed solution in comparison with native 5G deployments without any authorization. The 5G-WAVE integrated platform introduces 155ms authorization latency overhead to HTTP transactions in the 5G core service chains for a single slice. Additionally, the authorization latency increases linearly by 1.4 times with a 10-fold growth in the network. The 5G-WAVE framework enhances security by addressing OAuth 2.0 vulnerabilities and 3GPP key issues for network slicing security. The authors have provided public access to their code at~\cite{5gwavecode}.

 \label{sec:conc}
%%%%%%%%%%%%%%%%%%%%%%%%%%%%%%%%%

\section*{Acknowledgment}

We thank the anonymous reviewers for their valuable feedback. This work was supported in part by the U.S. Defense Advanced Research Projects Agency (DARPA) under agreement number HR001120C0155. The views, opinions, and findings contained in this article are those of the authors and should not be interpreted as representing the official views or policies, either expressed or implied, of the Defense Advanced Research Projects Agency or the Department of Defense.
%\newpage
\bibliographystyle{IEEEtran}
\bibliography{references.bib}

\end{document}